\newlength{\absize}
\documentclass[12pt]{article}
\usepackage{graphicx}
\usepackage{amssymb}
\usepackage{bbm}
\usepackage{braket}
\renewcommand{\arraystretch}{2}

\renewcommand{\arraystretch}{1.2}

\renewcommand{\bar}{\overline}

\newcommand{\spur}[1]{\!\not\! #1 \,}

\newcommand{\cL}{\mathcal{L}}

\newcommand{\cB}{\mathcal{B}}
\newcommand{\cC}{\mathcal{C}}
\newcommand{\cD}{\mathcal{D}}
\newcommand{\cE}{\mathcal{E}}

\newcommand{\cM}{\mathcal{M}}
\newcommand{\cN}{\mathcal{N}}

\newcommand{\pd}{\partial}

\newcommand{\be}{\begin{equation}}
\newcommand{\ee}{\end{equation}}
\newcommand{\bea}{\begin{eqnarray}}
\newcommand{\eea}{\end{eqnarray}}

\newcommand{\comment}[1]{}

\newcommand{\lra}{\mathop{\longrightarrow}\limits}
\newcommand{\sgn}{\mathop{\rm sgn}}

\begin{document}

\thispagestyle{empty}
\pagestyle{empty}
\newcommand{\starttext}{\newpage\normalsize
 \pagestyle{plain}
 \setlength{\baselineskip}{3ex}\par
 \setcounter{footnote}{0}
 \renewcommand{\thefootnote}{\arabic{footnote}}
 }
\newcommand{\preprint}[1]{\begin{flushright}
 \setlength{\baselineskip}{3ex}#1\end{flushright}}
\renewcommand{\title}[1]{\begin{center}\LARGE
 #1\end{center}\par}
\renewcommand{\author}[1]{\vspace{2ex}{\large\begin{center}
 \setlength{\baselineskip}{3ex}#1\par\end{center}}}
\renewcommand{\thanks}[1]{\footnote{#1}}
\renewcommand{\abstract}[1]{\vspace{2ex}\normalsize\begin{center}
 \centerline{\bf Abstract}\par\vspace{2ex}\parbox{\absize}{#1
 \setlength{\baselineskip}{2.5ex}\par}
 \end{center}}

\title{Gauge Boson Mass Dependence and
Chiral Anomalies in Generalized Massless Schwinger models}
\author{
 Howard~Georgi\thanks{\noindent \tt hgeorgi@fas.harvard.edu}
\\ \medskip
Center for the Fundamental Laws of Nature\\
Jefferson Physical Laboratory \\
Harvard University \\
Cambridge, MA 02138
 }
\date{\today}
\abstract{
I bosonize the position-space correlators of flavor-diagonal scalar fermion
bilinears in arbitrary generalizations of the Schwinger model with $n_F$
massless fermions coupled to $n_A$ gauge bosons for $n_F\geq n_A$.  For
$n_A=n_F$, the fermion bilinears can be bosonized in terms of $n_F$ scalars
with masses proportional to the gauge couplings.  As in the Schwinger
model, bosonization can be used to find all correlators, including those
that are forbidden in perturbation theory by anomalous chiral symmetries,
but there are subtleties when there is more than one gauge boson.  The new
result here is the general treatment of the dependence on gauge boson
masses in models with more than one gauge symmetry.  For $n_A<n_F$, there
are fermion bilinears with nontrivial anomalous dimensions and there are
unbroken chiral symmetries so some correlators vanish while others are
non-zero due to chiral anomlies.  Taking careful account of the dependence
on the masses, I show how the $n_A<n_F$ models emerge from $n_A=n_F$ as
gauge couplings (and thus gauge boson masses) go to zero.  When this is
done properly, the limit of zero gauge coupling is smooth. Our consistent
treatment of gauge boson masses guarantees that anomalous symmetries are
broken while unbroken chiral symmetries are preserved because correlators
that break the non-anomalous symmetries go to zero in the limit of zero
gauge coupling. 
}

\starttext

\section{Introduction\label{sec-intro}}

In this note, I analyze the
position-space correlators of flavor-diagonal scalar 
fermion bilinears in arbitrary generalizations of the massless
Schwinger model and focus on the behavior of gauge anomalies in the limit in
which some gauge couplings go to zero.   We will see that when the dimensional
gauge couplings are properly accounted for (which I believe is done here
for the first time) the limit is smooth.

We consider a general Schwinger model in 1+1 dimensions
with $n_F$ massless Dirac fermions, $\psi_\alpha$ for $\alpha=1$ to
$n_F$,
and $n_A\leq n_F$ vector bosons, $A^\mu_j$ for $j=1$ to $n_A$.

The Lagrangian is (using summation convention where it does not cause confusion)
\be
\cL_{Af} =
\left(\sum_{\alpha=1}^{n_F}\bar\psi_\alpha\,\left(i\spur\pd - e_{\alpha
j}\not\!\!A_j\right)
\,\psi_\alpha\right)
-\frac{1}{4}F_j^{\mu\nu} F_{j\mu\nu}
\label{vfn-Sommerfield-model2}
\ee
Note that we have assumed that the gauge couplings are diagonal in the
fermion space.  This is important to ensure that the model is exactly
solvable.  We could have non-diagonal couplings as long as the couplings to
different gauge bosons commute with one another.  But then we can
simultaneously diagonalize them with a unitary transformation on the
fermion fields, so we will assume that the gauge bosons only couple to
diagonal fermion currents.
The gauge couplings are an $n_F\times n_A$ matrix, $e$
in which
\begin{equation}
e_{\alpha j}\mbox{~is the coupling of the $j$th vector to the $\alpha$th
diagonal fermion current.}
\label{efa}
\end{equation}
If $n_F<n_A$, there are linear combinations of the gauge bosons that do not
couple to anything and can be safely ignored, so we will always assume
that $n_F\geq n_A$.

Each of the massless fermions generates a contribution to the vector boson
mass matrix~\cite{Lowenstein:1971fc} and we will use our freedom to redefine the
vector fields to diagonalize the physical vector boson mass
so we can write\footnote{This transformation may obscure charge
quantization.  See the discussion at the end section~\ref{sec-examples}.}
\begin{equation}
\sum_\alpha e_{\alpha j}e_{\alpha j'}=\pi\,m_j^2\,\delta_{jj'}
\quad\quad\quad
m_j^2=\frac{1}{\pi}\sum_\alpha e^2_{\alpha j}
\label{diagonal}
\end{equation}
Then the $m_j$ are the physical gauge boson masses.
The basis of the fermions is fixed up to unbroken flavor symmetries
because the currents must be diagonal.  
And unless some gauge boson
masses are equal, the basis of the gauge boson fields is also
fixed by (\ref{diagonal}). 

We work in Lorenz gauge, 
\begin{equation}
\partial_\mu A_j^\mu=0
\end{equation}

Then a generalization of the arguments of
\cite{Lowenstein:1971fc}\footnote{For details and definitions, see
\cite{Georgi:2022sdu}.  The conjecture in the paper was shown to be false
in \cite{Dempsey:2023gib} but we can still use the calculational tools.}
shows that the gauge invariant 
correlation functions to all orders in perturbation theory can be found using 
the free-field Lagrangian describing massless fermions, $\Psi_\alpha$,
massive bosons, $\cB_j$, and massless scalar ghosts, $\cC_j$
\be
\cL=\left(\sum_{\alpha=1}^{n_F}i\bar\Psi_\alpha\spur\pd\Psi_\alpha\right)
-\frac{m_j^2}{2}\cB_j^2+\frac{1}{2}\pd_\mu\cB_j\pd^\mu\cB_j
- \frac{1}{2}\pd_\mu\cC_j\pd^\mu\cC_j
\label{Sommerfield-redefined2}
\ee
using the replacements
\be
A_j^\mu = \epsilon^{\mu\nu}\pd_\nu\left(\cB_j-\cC_j\right)/m_j
\label{f-A-decomposition}
\ee
\be
\psi_\alpha = e^{-i\left(\pi\right)^{1/2}\,\left(e_{\alpha j}/m_j\right)\left(\cC_j-\cB_j\right)\gamma^5}\Psi_\alpha
\label{f-psi-redef-vcb}
\ee

The massive and the massless free scalar propagators, respectively,
are\footnote{$K_0$ is the modified Bessel function of the second kind.} 
\be
-i\braket{0|T \cB_j\left(x\right) \cB_k\left(0\right)|0}=\delta_{jk}
\int\frac{d^2p}{\left(2\pi\right)^2}\frac{e^{-ipx}}{p^2 - m_j^2 +
i\epsilon} 
= -\delta_{jk}\frac{i}{2\pi}K_0\left(m_j\sqrt{-x^2 + i\epsilon}\right)
\label{bprop}
\ee
\be
-i\braket{0|T \cC_j\left(x\right) \cC_k\left(0\right)|0}=
- \delta_{jk}\int\frac{d^2p}{\left(2\pi\right)^2}\frac{e^{-ipx}}{p^2 + i\epsilon} 
= -\delta_{jk}\frac{i}{2\pi}\ln\left(\xi m_j\sqrt{-x^2+i\epsilon}\right) 
\label{cprop}
\ee
\be
\xi\equiv e^{\gamma_E}/2\mbox{~~where $\gamma_E$ is Euler's constant}
\label{xi}
\ee
The arbitrary dimensional constant in the logarithmic ghost propagators has
been fixed so that the $\cB_j$ and $\cC_j$ propagators exactly cancel when
the gauge couplings vanish, $m_j=0$. 

We will focus on the position-space correlators of the
flavor-diagonal fermion-bilinear scalar
operators
\be
{\renewcommand{\arraystretch}{1.7}\begin{array}{l}
\displaystyle
O_{\alpha+}\equiv \psi_{2\alpha}^*\,\psi_{1\alpha} 
= e^{-2i \sum_j\left(e_{\alpha j}/m_j\right)\left(\cB_j-\cC_j\right)}
\Psi^*_{2\alpha}\Psi_{1\alpha}
\\
\displaystyle
O_{\alpha-}\equiv \psi_{1\alpha}^*\,\psi_{2\alpha} 
= e^{2i \sum_j\left(e_{\alpha j}/m_j\right)\left(\cB_j-\cC_j\right)}
\Psi^*_{1\alpha}\Psi_{2\alpha}=O_{\alpha+}^*
\end{array}}
\label{oalphaf}
\ee

While (\ref{oalphaf}) gives a complete description of the $O_{\alpha\pm}$
correlators to all orders in perturbation theory, the nonperturbative
effects of gauge anomalies are much more transparent if we bosonize.  For
massless free fermions in 1+1, \textbf{any non-zero} correlator of diagonal fermion
bilinears can be calculated by 
replacing the bilinears with exponentials of free massless scalar fields
according to  
\begin{equation}
\Psi^*_{2\alpha}\Psi_{1\alpha}\to
 \frac{\xi \cM_\alpha}{2\pi}\,e^{-2i\pi^{1/2}\cD^\alpha}
\quad\quad
\Psi^*_{1\alpha}\Psi_{2\alpha}\to
 \frac{\xi \cM_\alpha}{2\pi}\,e^{2i\pi^{1/2}\cD^\alpha}
\label{bosonize}
\end{equation}
The $\cM_\alpha$s are arbitrary and correlated with the propagators of the
bosonization fields, $\cD^\alpha$.
\be
-i\braket{0|T \cD^\alpha\left(x\right) \cD^\beta\left(0\right)|0}=\delta^{\alpha\beta}
\int\frac{d^2p}{\left(2\pi\right)^2}\frac{e^{-ipx}}{p^2 + i\epsilon} 
= \delta^{\alpha\beta}\frac{i}{2\pi}\ln\left(\xi \cM_\alpha\sqrt{-x^2+i\epsilon}\right) 
\label{dprop}
\ee
There is no mass in the position space correlators of the
fermion bilinears and the mass $\cM_\alpha$ is introduced by the
bosonization procedure 
which requires a mass to get the dimensions right. 
(\ref{bosonize}) is a straightforward consequence of Fermi-Dirac statistics.  
Then
\be
O_{\alpha\pm} 
= \frac{\xi \cM_\alpha}{2\pi}\,
e^{\mp2i \sum_{j=1}^{n_A}\left(e_{\alpha j}/m_j\right)\left(\cB_j-\cC_j\right)}
\,e^{\mp2i\pi^{1/2}\cD^\alpha}
\label{oalpha}
\ee

The key to using (\ref{oalpha}) most effectively will be to cancel the
ghost fields with linear combinations of the bosonization fields.  This will be easier if
we change notation slightly.

First, we define an orthogonal $n_F\times n_F$ matrix $\eta_{\alpha j}$
where the $\alpha$ index labels the fermion as usual but the $j$ index is
extended to make the matrix square and a subset $b$ of the $j$ indices are
associated with the gauge fields.
\begin{equation}
\eta\,\eta^T=I\quad\quad
\eta_{\alpha j}=\frac{e_{\alpha j}}{m_j\sqrt{\pi}}
\mbox{~~for~~$j\in b$}
\label{xija}
\end{equation}
we can write
\be
O_{\alpha \pm} 
= \frac{\xi \cM_\alpha}{2\pi}\,
e^{\mp 2i\pi^{1/2} \sum_{j\in b}\eta_{\alpha j}\left(\cB_j-\cC_j\right)}
\,e^{\mp2i\pi^{1/2}\cD^\alpha}
\label{oalphaxi}
\ee

Now we can define linear combinations of the bosonization fields
\begin{equation}
 \cD_j\equiv \eta_{\alpha j}
\cD^\alpha 
\end{equation}
and write
\be
O_{\alpha\pm} 
= \frac{\xi \cM_\alpha}{2\pi}\,
e^{\mp2i\pi^{1/2}\sum_{j\in b} \eta_{\alpha j}\left(\cB_j-\cC_j\right)}
\,e^{\mp2i\pi^{1/2}\sum_{j}\eta_{\alpha j} \cD_j}
\label{oalphaxichi}
\ee

In this form, it looks like we can cancel the $\cD_j$ fields for $j\in b$
with the ghosts.  That is precisely what happens in the original 
Schwinger model and in the Schwinger model with flavors.  But in general,
this cancellation is not exact because of the different masses associated
with different gauge couplings.

The $\cD_j$ propagator is
\be
-i\braket{0|T \cD_j\left(x\right) \cD_k\left(0\right)|0}=
\sum_\alpha\eta_{j\alpha}\eta_{k\alpha}
\frac{i}{2\pi}\ln\left(\xi \cM_\alpha\sqrt{-x^2+i\epsilon}\right) 
\label{djprop}
\ee
Thus for the $\cD_j$ for $j=1$ to $n_A$ to exactly
cancel the ghosts $\cC_j$
we \textbf{must} have 
\begin{equation}
\sum_\alpha\eta_{\alpha j}\eta_{\alpha k}\log\cM_\alpha=\delta_{jk}\log
m_j
\mbox{~~~for $j$ or $k\in b$}
\label{must}
\end{equation}
and (\ref{oalphaxichi}) 
would be most useful if the $\cD_j$ propagator were diagonal for
all $j$.  In general, this is impossible and the most useful thing we can do is to
choose a common bosonization mass $\cM_\alpha=m$ for the $\cD^\alpha$s:
\begin{equation}
\cM_\alpha=m\mbox{~~~for all $\alpha$.}
\label{all}
\end{equation}
so that
\be
-i\braket{0|T \cD_j\left(x\right) \cD_k\left(0\right)|0}=
\delta_{jk}\frac{i}{2\pi}\ln\left(\xi m\sqrt{-x^2+i\epsilon}\right) 
\label{djpropm}
\ee
We could always cancel the ghost for a single $j$ by choosing $m=m_j$.  This is
the way the issue is typically handled in models with a single gauge
boson but it is not adequate for $n_A>1$.  

If we adopt (\ref{djpropm}), we can \textbf{almost} cancel the ghosts in general.
\be
O_{\alpha\pm} 
= \frac{\xi m}{2\pi}\,
e^{\mp2i\pi^{1/2}\sum_{j\in b} \eta_{\alpha j}\left(\cB_j-\cE_j\right)}
\,e^{\mp2i\pi^{1/2}\sum_{j\in b'}\eta_{\alpha j} \cD_j}
\label{oalphaebd}
\ee
where $b'$ is the complement of $b$ and the $\cE_j$ are ``constant'' fields
\begin{equation}
\cE_j=\cC_j-\cD_j\mbox{~~~for $j\in b$}
\end{equation}
with propagators
\be
-i\braket{0|T \cE_j\left(x\right) \cE_k\left(0\right)|0}=
\delta_{jk}\frac{i}{2\pi}\log\left( m-m_j\right) 
\label{ejpropm}
\ee
so the Wick expansion gives
\be
\braket{0|T e^{2i\pi^{1/2}\cE_j\left(x\right)}\,e^{\mp 2i\pi^{1/2} \cE_j\left(0\right)}|0}=
\left(m_j/m\right)^{\pm2}
\label{ejpropme}
\ee

The $\cE_j$ fields keep the engineering dimensions right while
eliminating the dependence on the arbitrary bosonization mass $m$.
(\ref{oalphaebd}-\ref{ejpropm}) 
can be used to calculate any matrix element that is non-zero in
perturbation theory and we can simplify
(\ref{oalphaebd}) further by eliminating the $\cE_j$
operators.\footnote{Note that we will soon use (\ref{oalphaebd}) to
calculate other matrix elements, but because these more general results
follow from the perturbative calculation and cluster decomposition, the
simplification will work in general.}

Look at a general perturbatively non-zero 
correlator involving only bilinears (no higher powers).   
If there are $n_\alpha$ $O_{\alpha+}$s 
there must also be $n_\alpha$ $O_{\alpha-}$s or the correlator would vanish in
perturbation theory.  We can look separately at the contributions
from each $\cE_j$ because the propagators don't mix.  So we are interested
in 
\be
\braket{0|T \prod_{r=1}^{n_\alpha}e^{2i\pi^{1/2}\eta_{\alpha
j}\cE_j\left(x_r\right)}\,
e^{-2i\pi^{1/2}\eta_{\alpha
j}\cE_j\left(y_r\right)}|0}
\label{nalphaej}
\ee
We have put arguments $\left(x_r\right)$ and $\left(y_r\right)$ on the $\cE_j$
from $O_{\alpha+}\left(x_r\right)$ and  $O_{\alpha-}\left(y_r\right)$ for
notational convenience, but
the $\cE_j$ are constant operators so the Wick expansion of
(\ref{nalphaej}) is independent of the coordinates.  The contribution from
contractions of the 
$\cE_j\left(x_r\right)$s with the $\cE_j\left(y_r\right)$s is 
\begin{equation}
\left(\frac{m_j^2}{m^2}\right)^{n_\alpha^2\eta_{\alpha j}^2}
\end{equation} 
The contribution from contractions of the
$\cE_j\left(x_r\right)$s with $\cE_j\left(x'_r\right)$s is 
\begin{equation}
\left(\frac{m_j^2}{m^2}\right)^{-\left(n_\alpha^2-n_\alpha\right)\eta_{\alpha j}^2/2}
\end{equation} 
as is the
contribution from contractions of the
$\cE_j\left(y_r\right)$s with $\cE_j\left(y'_r\right)$s.  So the total contribution is
\begin{equation}
\left(\frac{m_j^2}{m^2}\right)^{n_\alpha\eta_{\alpha j}^2}
\end{equation} 
This means that we can associate a factor
\begin{equation}
N_\alpha=\prod_{j\in b}\left(\frac{m_j}{m}\right)^{\eta_{\alpha j}^2}
\label{nalpha}
\end{equation} 
with each operator in each $O_{\alpha+}$-$O_{\alpha-}$ pair and completely
capture the effects of the $\cE_j$s for perturbatively non-zero correlators.

Now we can eliminate the $\cE_j$s and write 
\be
{\renewcommand{\arraystretch}{2.5}\begin{array}{c}
\displaystyle
O_{\alpha\pm} 
= \frac{\xi m}{2\pi}\,
e^{2\pm i\pi^{1/2}\sum_{j\in b} \eta_{\alpha j}\left(\cE_j-\cB_j\right)}
\,e^{2\mp i\pi^{1/2}\sum_{j\in b'}\eta_{\alpha j} \cD_j}\to\\
\displaystyle
\frac{\xi m}{2\pi}\,
N_\alpha\,e^{-2\pm i\pi^{1/2}\left(\sum_{j\in b} \eta_{\alpha j}\cB_j
+\sum_{j\in b'}\eta_{\alpha j} \cD_j\right)}
\end{array}}
\label{oalphabdmj}
\ee

We can use (\ref{oalphabdmj}) to calculate any correlator that is non-zero
in perturbation theory.  A particularly nice feature of (\ref{oalphabdmj})
is the way it encodes the diagonal chiral symmetries.  The theory is invariant under
global translations of the massless $\cD_j$ fields
for $j\in b'$.  These translations generate the $n_F-n_A$ unbroken chiral symmetries
on the fermion bilinears.  
\begin{equation}
\mbox{$O_{\alpha\pm}$ has $j$ chiral charge $q_j=\pm\eta_{\alpha j}$ for
$\delta\cD_j=\theta_j$ with $j\in b'$.}
\label{chiral-charge}
\end{equation} 
There are no unbroken chiral transformations from translations of $\cD_j$
for $j\in b$, because these $\cD_j$s have been eaten by the ghosts.
The corresponding chiral transformation from translations of the $\cB_j$
for $j\in b$ are softly broken by 
the $\cB_j$ mass terms generated by the gauge anomalies.

\section{Cluster Decomposition}

While we only calculate to all orders in perturbation theory, in some
situations, cluster decomposition
gives us nonperturbative information.
For any combination of $n_o$ fermion bilinears in some region of
space-time, we can 
look at a correlator that also includes all the conjugate fields in a
region far away
in space as in
\begin{equation}
\Braket{0|T\,\prod_{u=1}^{n_o} O_{\alpha_us_u}(x_{u})\,
O_{\alpha_{u}s_{u}}^*(z+y_{u})|0}
\label{oxyz}
\end{equation}
where the $s_u$ are $\pm1$, as in (\ref{oalphaf}) and the $x_a$ and
$y_a$ are clustered in some region of size $\ell$ ($-(x_a-x_b)^2<\ell^2$)
around the (arbitrary) origin and $z$ is
a large space-like 2-vector.
This correlator is calculable in perturbation theory. Using
(\ref{oalphabdmj}), we can write it as
\begin{equation}
\left(\frac{\xi m}{2\pi}\right)^{2n_o}\left(\prod_u
N_{\alpha_u}\right)\,\left(X\,Y\,Z\right)
\label{oxyz1}
\end{equation}
where
\begin{equation}
{\renewcommand{\arraystretch}{2}\begin{array}{c}
X=\prod_{u\neq u'}\exp\left[-2s_us_{u'}\sum_j\eta_{\alpha_{u} j}\eta_{\alpha_{u'}j}
\Delta_j(x_u-x_{u'})\right]
\\
Y=\prod_{u\neq u'}\exp\left[-2s_us_{u'}\sum_j\eta_{\alpha_{u} j}\eta_{\alpha_{u'}j}
\Delta_j(y_u-y_{u'})\right]
\\
Z=\prod_{u, u'}\exp\left[2s_us_{u'}\sum_j\eta_{\alpha_{u} j}\eta_{\alpha_{u'}j}
\Delta_j(z+y_{u'}-x_u)\right]
\end{array}}
\label{oxyz2}
\end{equation}
where
\begin{equation}
\Delta_j(x)=
\left\{
{\renewcommand{\arraystretch}{1.7}\begin{array}{ll}
K_0\left(m_j\sqrt{-x^2+i\epsilon}\right)&\mbox{for $j\in b$}\\
-\log\left(m\sqrt{-x^2+i\epsilon}\right)&\mbox{for $j\in b'$}
\end{array}}
\right.
\label{oxyz3}
\end{equation}
We can now study the correlator as $-z^2$ goes to infinity.

If $b'\neq\emptyset$, that is for $n_F>n_A$ 
the correlator will in general go to zero, with the
$Z$ factor falling off like
\begin{equation}
\frac{1}{\left(m\sqrt{-x^2+i\epsilon}\right)^{2\sum_{j\in b'}Q_j^2}}
\end{equation}
where $Q_j$ is the total $j$ chiral charge of the operators in the $X$
region,
\begin{equation}
Q_j=\sum_{u}s_u\eta_{\alpha_{u} j}
\end{equation}
Thus $\sum_{j\in b'}Q_j^2$ is the anomalous dimension of the leading
operators in the operator product expansion of the set
$\{O_{\alpha_us_u}(x_u)\}$ and I will refer to this as the anomalous dimension
of the set and we will have more to say about it below.

But we will begin by assuming that
$n_A=n_F$ so there are no massless fields
($b'=\emptyset$) and
\begin{equation}
\Delta_j(x)=K_0\left(m_j\sqrt{-x^2+i\epsilon}\right) \;\forall\;j
\end{equation}
Then for
sufficiently large $z$, all the Bessel functions go to $0$ and $Z\to1$.
If the
correlator is non-zero as the
regions move infinitely far apart, it must factor into a product of contributions 
in the two regions.  
\begin{equation}
\Braket{0|T\,\prod_{u=1}^{n_o} O_{\alpha_us_u}(x_{u})\,
O_{\alpha_{u}s_{u}}^*(z+y_{u})|0}
\lra_{-z^2\to\infty}
\left(\frac{\xi m}{2\pi}\right)^{2n_o}\left(\prod_u N_{\alpha_u}\right)\,X\,Y
\label{oxyzo}
\end{equation}
This gives non-perturbative information about anomalies.
Conversely, if the correlators in the two regions are forbidden by unbroken
symmetries, the correlator (\ref{oxyz}) must vanish as $z\to\infty$.

The above argument means that for $n_A=n_F$, up to
phases~\cite{Jayewardena:1988td,Sachs:1991en,Smilga:1992hx} that are
arbitrary and that we will set to zero, 
we can use (\ref{oxyz}-\ref{oxyz3}) to
calculate any correlator using (\ref{oalphabdmj}) 
whether or not it is non-zero in perturbation theory and the result is
\begin{equation}
{\renewcommand{\arraystretch}{2.5}\begin{array}{c}
\displaystyle
\Braket{0|T\,\prod_{u=1}^{n_o} O_{\alpha_us_u}(x_{u})|0}
=\left(\frac{\xi m}{2\pi}\right)^{n_o}\left(\prod_u N_{\alpha_u}\right)
\\
\displaystyle
\prod_{u\neq u'}\exp\left[-2s_us_{u'}\sum_j\eta_{\alpha_{u} j}\eta_{\alpha_{u'}j}
K_0\left(m_j\sqrt{-(x_u-x_{u'})^2+i\epsilon}\right)\right]
\end{array}}
\label{x}
\end{equation}
This makes sense because when $b'=\emptyset$ with no $\cD$ fields, there are
no global chiral symmetries 
that are not broken by the gauge boson mass terms.  So there are no
constraints on the use of (\ref{oxyz}-\ref{oxyz3}). 
Because the bosonization mass $m$ is arbitrary, it will cancel explicitly
in (\ref{x}), so we can also write (\ref{x})
entirely in terms of the $m_j$s as
\begin{equation}
{\renewcommand{\arraystretch}{2.5}\begin{array}{c}
\displaystyle
\Braket{0|T\,\prod_{u=1}^{n_o} O_{\alpha_us_u}(x_{u})|0}
=\left(\frac{\xi}{2\pi}\right)^{n_o}\left(\prod_u M_{\alpha_u}\right)
\\
\displaystyle
\prod_{u\neq u'}\exp\left[-2s_us_{u'}\sum_j\eta_{\alpha_{u} j}\eta_{\alpha_{u'}j}
K_0\left(m_j\sqrt{-(x_u-x_{u'})^2+i\epsilon}\right)\right]
\end{array}}
\label{xnom}
\end{equation}
where 
\begin{equation}
M_\alpha=\prod_{j\in b}m_j^{\eta_{\alpha j}^2}
\label{malpha}
\end{equation} 

Equations (\ref{nalpha}), (\ref{x}), (\ref{malpha}), and (\ref{xnom}) are the basic
technical results of this note. (\ref{nalpha},\ref{x}) and
(\ref{malpha},\ref{xnom}) are equivalent, but  (\ref{nalpha},\ref{x}) 
is more convenient
for taking $m_j$s to zero which we will do in the next section. The $m_j$
dependence in (\ref{nalpha}) and 
(\ref{malpha}) is useful in understanding the structure of the general theory.

Now we can use (\ref{nalpha},\ref{x}) to analyze the
general theory for $b'\neq\emptyset$.  While we could do the general analysis
directly, constructing the general theory as
the limit
of (\ref{x}) as some gauge couplings go to zero automatically includes the
nonperturbative effects of the gauge anomalies and will give us
insights into their structure. 
It may seem surprising that this works at all because the corresponding
limit in 3+1 dimensions is quite dangerous.  But here everything goes
smoothly because we have properly included the important dimensional
parameters in (\ref{nalpha}).  

In particular, look at the limit $m_j\to0$ for some $j$ in (\ref{x}).
Because of the product structure of the correlator, we can focus just on
the factors that depend on $j$:
\begin{equation}
\left(\prod_u \left(\frac{m_j}{m}\right)^{\eta_{\alpha_uj}^2}\right)
\prod_{u\neq u'}\exp\left[-2s_us_{u'}\eta_{\alpha_{u} j}\eta_{\alpha_{u'}j}
K_0\left(m_j\sqrt{-(x_u-x_{u'})^2+i\epsilon}\right)\right]
\label{xj1}
\end{equation}
If $m_j\sqrt{-(x_u-x_{u'})^2}$ is much less than $1$ for all the coordinate
pairs, we can approximate the Bessel functions by logs: 
\begin{equation}
\left(\prod_u \left(\frac{m_j}{m}\right)^{\eta_{\alpha_uj}^2}\right)
\prod_{u\neq u'}\exp\left[2s_us_{u'}\eta_{\alpha_{u} j}\eta_{\alpha_{u'}j}
\log\left(m_j\sqrt{-(x_u-x_{u'})^2+i\epsilon}\right)\right]
\label{xj2}
\end{equation}
This can be written
\begin{equation}
\left(\frac{m_j}{m}\right)^{\left(\sum_u\eta_{\alpha_uj}^2\right)
+\left(\sum_{u\neq u'}2s_us_{u'}\eta_{\alpha_{u} j}\eta_{\alpha_{u'}j}\right)}
\prod_{u\neq u'}\exp\left[2s_us_{u'}\eta_{\alpha_{u} j}\eta_{\alpha_{u'}j}
\log\left(m\sqrt{-(x_u-x_{u'})^2+i\epsilon}\right)\right]
\label{xj3}
\end{equation}
\begin{equation}
=\left(\frac{m_j}{m}\right)^{
\left(\sum_{u}s_u\eta_{\alpha_{u} j}\right)^2}
\prod_{u\neq u'}\exp\left[2s_us_{u'}\eta_{\alpha_{u} j}\eta_{\alpha_{u'}j}
\log\left(m\sqrt{-(x_u-x_{u'})^2+i\epsilon}\right)\right]
\label{xj4}
\end{equation}
Now we can take the limit $m_j\to0$.  The limit vanishes unless
\begin{equation}
Q_j=\sum_{u}s_u\eta_{\alpha_{u} j}=0
\label{chiral-charge0j}
\end{equation}
This is expected from (\ref{chiral-charge}). 
In
(\ref{chiral-charge0j}), $Q_j$ 
is the total $j$ chiral charge of the product of operators in the
correlator. 
The limit is just the statement that the correlator vanishes
if the product of operators carries an unbroken chiral symmetry.
But (\ref{xj4}) shows us exactly how the limit is approached.

If $Q_j$ vanishes, the first factor in
(\ref{xj4}) is $1$,  the $m_j$
dependence disappears, 
and the bosonization mass $m$ shows up in the second factor in the expected
way for a bosonization field.  Notice that it was critical here to include
the $N(\alpha)$ factor.  

We can now find the correlators for a general model by starting with all
gauge couplings non-zero and then taking $g_j\to0$ for $j\in b'$, so in
general the collections of $n_o$ operators that have non-zero correlators will satisfy
\begin{equation}
Q_j=\sum_{u}s_u\eta_{\alpha_{u} j}=0\quad\forall\,j\in b'
\label{chiral-charge0}
\end{equation}
The correlators are then simply
\begin{equation}
{\renewcommand{\arraystretch}{2.5}\begin{array}{c}
\displaystyle
\Braket{0|T\,\prod_{u=1}^{n_o} O_{\alpha_us_u}(x_{u})|0}
=\left(\frac{\xi m}{2\pi}\right)^{n_o}\left(\prod_u N_{\alpha_u}\right)
\\
\displaystyle
\prod_{u\neq u'}\exp\left[-2s_us_{u'}\sum_j\eta_{\alpha_{u} j}\eta_{\alpha_{u'}j}
\Delta_j\left(x_u-x_{u'}\right)\right]
\end{array}}
\label{xmj0}
\end{equation}
where as in (\ref{oxyz3})
\begin{equation}
\Delta_j(x)=
\left\{
{\renewcommand{\arraystretch}{1.7}\begin{array}{ll}
K_0\left(m_j\sqrt{-x^2+i\epsilon}\right)&\mbox{for $j\in b$}\\
-\log\left(m\sqrt{-x^2+i\epsilon}\right)&\mbox{for $j\in b'$}
\end{array}}
\right.
\label{delta}
\end{equation}

\section{Anomalies}

 Obviously, as $m_j\propto g_j\to0$, the gauge anomalies associated with
$A^\mu_j$ disappear.  But the requirement of (\ref{chiral-charge0}) puts
interesting constraints on the form of the remaining anomalies for $m_j=0$.
In general a set of operators $O_{\alpha_us_u}$ satisfying (\ref{chiral-charge0})
can be decomposed into irreducible sets which cannot be split into subsets
satisfying (\ref{chiral-charge0}).  These always include any single
operators $\{O_{\alpha+}\}$ and $\{O_{\alpha-}\}$ for which 
$\eta_{\alpha j}=0\,\forall j\in b'$  and the $\pm$ pairs, 
$\{O_{\alpha+},O_{\alpha-}\}$ for each $\alpha$ for which $\eta_{\alpha
j}\neq0$ for any $j\in b'$.  I will refer to any other irreducible sets as ``anomaly sets''
because they are related to the remaining gauge anomalies.  They are
characterized by a set of integers ${n_\alpha}$:
\begin{equation}
\mbox{$S({n_\alpha})$ consists of $|n_\alpha|$ copies of
$O_{\alpha\sgn(n_\alpha)}$ for each $\alpha$}
\end{equation}
In terms of the $n_\alpha$, (\ref{chiral-charge0}) is
\begin{equation}
\sum_\alpha |n_\alpha|\sgn(n_{\alpha})\eta_{\alpha j}
=\sum_\alpha n_\alpha\eta_{\alpha j}
=0\quad\forall\,j\in b'
\label{qjn}
\end{equation}
Then (\ref{xija}) implies that $n_\alpha$ is a linear combination of
$\eta_{\alpha j}$ for $j\in b$,
\begin{equation}
n_\alpha=\sum_{j\in b}\cN_j\eta_{\alpha j}
\label{nb}
\end{equation}
It makes sense that the irreducible anomaly sets depend only on the
remaining gauge couplings.

\section{Examples\label{sec-examples}}

If there is only one gauge boson, there is only one $\eta_{\alpha j}$ for
$j\in b$ --- call it $\eta_{\alpha1}$.  Then (\ref{nb}) is
\begin{equation}
n_\alpha\propto\eta_{\alpha1}
\label{nb1}
\end{equation}
which fixes the irreducible
anomaly sets (if they exist) up to a sign (which is just associated with 
complex conjugation of all the
bilinears).  But $n_\alpha$ exists if an only if 
the components of $\eta_{\alpha1}$ are commensurate.

Here is the 2-flavor Schwinger model with $b=\{1\}$:
\begin{equation}
\eta_{\alpha 1}=\left(\begin{array}{c}
1/\sqrt{2}\\1/\sqrt{2}
\end{array}\right)
\quad
\eta_{\alpha 2}=\left(\begin{array}{c}
-1/\sqrt{2}\\1/\sqrt{2}
\end{array}\right)
\quad
n_\alpha=\left(\begin{array}{c}
\pm1\\ \pm1
\end{array}\right)
\label{s2n}
\end{equation}
The irreducible anomaly sets are $\{O_{1\pm},O_{2\pm}\}$.

Here is a similar model again with $b=\{1\}$ but 
gauge couplings that differ by a factor of 2.
\begin{equation}
\eta_{\alpha 1}=\left(\begin{array}{c}
2/\sqrt{5}\\1/\sqrt{5}
\end{array}\right)
\quad
\eta_{\alpha 2}=\left(\begin{array}{c}
-1/\sqrt{5}\\2/\sqrt{5}
\end{array}\right)
\quad
n_\alpha=\left(\begin{array}{c}
\pm2\\ \pm1
\end{array}\right)
\label{s25n}
\end{equation}
The irreducible anomaly sets are $\{2\times O_{1\pm},O_{2\pm}\}$.

Again with $b=\{1\}$
\begin{equation}
\eta_{\alpha 1}=\left(\begin{array}{c}
n_1/\sqrt{n_1^2+n_2^2}\\n_2/\sqrt{n_1^2+n_2^2}
\end{array}\right)
\quad
\eta_{\alpha 2}=\left(\begin{array}{c}
-n_2/\sqrt{n_1^2+n_2^2}\\n_1/\sqrt{n_1^2+n_2^2}
\end{array}\right)
\label{sn1n2}
\end{equation}
Now if $n_1$ and $n_2$ are relatively prime, 
\begin{equation}
n_\alpha=\left(\begin{array}{c}
\pm n_1\\ \pm n_2
\end{array}\right)
\label{sn1n2n}
\end{equation}
and the irreducible anomaly sets are $\{n_1\times O_{1\pm},n_2\times O_{2\pm}\}$.
If $n_1$ and $n_2$ have a common factor $n_c$ the irreducible
anomaly sets are $\{(n_1/n_c)\times O_{1\pm},(n_2/n_c)\times O_{2\pm}\}$. 

Finally for $b=\{1\}$, if the gauge couplings are not commensurate  --- as in.
\begin{equation}
\eta_{\alpha 1}=\left(\begin{array}{c}
\sqrt{2/3}\\ \sqrt{1/3}
\end{array}\right)
\quad
\eta_{\alpha 2}=\left(\begin{array}{c}
-\sqrt{1/3}\\ \sqrt{2/3}
\end{array}\right)
\label{sn}
\end{equation}
there are no
anomaly sets.~\cite{Hosotani:1998za}

If there is more than gauge boson, more complicated scenarios are possible.  If the
components of $\eta_{\alpha j}$ are commensurate
for each $j\in b$, there is an independent
pair for anomaly sets for each $j\in b$ and they may be reducible.  
For example, with two gauge
bosons, $b={1,2}$, we could have a ``diagonal color'' 
model:~\cite{Belvedere:1978fj,Steinhardt:1980rz,GamboaSaravi:1981zd,Belvedere:1986ik,Georgi:2019tch}
\begin{equation}
\eta_{\alpha1}=\left(
\begin{array}{c}
1/\sqrt{2}\\-1/\sqrt{2}\\0
\end{array}\right)
\quad
\eta_{\alpha2}=\left(
\begin{array}{c}
1/\sqrt{6}\\1/\sqrt{6}\\-2/\sqrt{6}
\end{array}\right)
\quad
\eta_{\alpha3}=\left(
\begin{array}{c}
1/\sqrt{3}\\1/\sqrt{3}\\1/\sqrt{3}
\end{array}\right)
\end{equation} 
The irreducible anomaly sets are 
$\{O_{1\pm},O_{2\mp}\}$,
$\{O_{2\pm},O_{3\mp}\}$, and
$\{O_{3\pm},O_{1\mp}\}$.

But incommensurate charges can further eliminate all the anomaly sets as in
(\ref{sn}) or just limit them, as in this example (again with $b={1,2}$):
\begin{equation}
\eta_{\alpha1}=\left(
\begin{array}{c}
 1/2 \\
 \left(2+\sqrt{2}\right)/4 \\
 \left(2-\sqrt{2}\right)/4
 \end{array}
\right)
\quad
\eta_{\alpha2}=\left(
\begin{array}{c}
 -1/2 \\
 \left(2-\sqrt{2}\right)/4 \\
 \left(2+\sqrt{2}\right)/4
 \end{array}
\right)
\quad
\eta_{\alpha3}=\left(
\begin{array}{c}
 1/\sqrt{2} \\
-1/2\\ 1/2
 \end{array}
\right)
\label{contrived}
\end{equation}
Here because the only linear combination of $\eta_{\alpha1}$ and
$\eta_{\alpha2}$ with commensurate components is
$\eta_{\alpha1}+\eta_{\alpha2}$, the only irreducible anomaly sets are
$\{O_{2\pm},O_{3\pm}\}$. 

One might ask\footnote{as a clever referee did} whether commensurate
components of the $\eta_{\alpha j}$ vectors is associated with quantization
of charge.  That is obviously the case for $n_A=1$, as shown in 
examples (\ref{s2n},\ref{s25n},\ref{sn}).  
For $n_A>1$, the
situation is more complicated because generically, the mass-eigenstates of
the gauge fields will be linear combinations of the fields in the original
Lagrangian.  But the $\eta_{\alpha j}$ for $j\in b'$ are orthogonal to ALL
of the $\eta_{\alpha j}$ for $j\in b$ and thus to any linear combination.
So if there is any linear combination of the $\eta_{\alpha j}$ for $j\in b$
with commensurate coefficients, it will be associated with an anomaly set
AND the corresponding combination of gauge fields will be coupled to a
quantized charge.  For example, in (\ref{contrived}), the combination
$\eta_{\alpha1}+\eta_{\alpha 2}$ describes a coupling to a quantized charge.

In general, the independent anomaly sets will be
associated with the 
independent linear combinations coupled to quantized charges.  This is
exactly what we would expect from Coleman's interpretation of a $\theta$
parameter as a background electric field~\cite{Coleman:1976uz}.  Only the
linear combinations 
corresponding to quantized charges will have $\theta$ parameters because a
background field coupled to a non-quantized charge can be canceled by pair
production of massless fermions.  Thus the independent anomaly sets will be
associated with the independent $\theta$ parameters.  This connection can
be obscured in models with more than one gauge boson by a nontrivial gauge
boson mass matrix, but we have shown how a complete analysis of the mass
dependence describes the general case.

\section{Conclusions\label{sec:end}}

While the physics is trivial, the correlators in
generalizations of the massless Schwinger model
depend in nontrivial ways on the gauge boson masses. I believe that the
complete analysis of the mass dependence in this note clarifies the
relationship between models with different numbers of gauge bosons and
chiral anomalies.

\section{Acknowledgements}
I am grateful to Igor Klebanov for helpful comments.  
This project has received support from the European Union's
Horizon 2020 research and innovation programme under the Marie 
Skodowska-Curie grant agreement No 860881-HIDDeN.

\bibliography{up4}\end{document}